\documentclass[aps,prl,reprint,amsmath,amssymb]
{revtex4-1}
\usepackage{graphicx}

\begin{document}
\title{Evidence for superconductivity in Rb metal above 55 GPa pressure}
\author{Yuhang Deng}
\affiliation{Department of Physics, Washington University, St Louis, Missouri, USA}
\author{James S. Schilling}
\affiliation{Department of Physics, Washington University, St Louis, Missouri, USA}

\date{\today}

\begin{abstract}
The only alkali metal known to be superconducting at ambient pressure is Li at
0.4 mK. Under 30 GPa pressure \textit{T}$_{c}$ for Li rises to 14 K. In
addition, nearly 50 years ago the heavy alkali metal Cs was reported to become
superconducting near 1.3 K at 12 GPa. In the present experiment the superconductivity
of Cs under pressure is confirmed. In addition, strong evidence is presented in electrical
resistivity measurements that neighboring Rb also becomes superconducting near
2 K at 55 GPa as it enters the \textit{oC}16 phase, as for Cs, where
\textit{T}$_{c}$ decreases under the application of pressure. It would seem
likely that under the right temperature/pressure conditions all alkali metals,
including metallic hydrogen, will join the ranks of the superconducting
elements. With the addition of Rb, 55 of the 92 naturally occurring elements
are superconducting at ambient or high pressure.
\end{abstract}

\maketitle

At first glance the alkali metals would appear to be rather mundane compared
to the other elemental solids in the periodic table: none of the alkalis are
magnetic and only one, Li, is known to be superconducting at ambient pressure,
but then only at the exceedingly low temperature of $T_{\text{c}}\simeq$ 0.4
mK \cite{tuoriniemi}. Their simple bcc structure and free-electron character
make the alkalis ideal simple metals with very nearly spherical Fermi
surfaces. As for all monovalent metals, their low electronic density of states
favors weak electron-phonon interactions that are insufficient, except for Li,
to support superconductivity \cite{schilling}.

The large atomic volume and very high compressibility of all alkalis lead to
significant changes in their properties under the application of very high
pressures. The atomic volume of Cs metal shrinks five-fold under $\sim$\ 50
GPa \cite{takemura}. As a result, the once free-electron alkalis with cubic
bcc crystal structure become \textit{electrides} under sufficient pressure
where the conduction electrons no longer surround each cation but are forced
into interstitial lattice sites \cite{neaton}, becoming \textit{de facto} anions
\cite{dye}. In addition, the marked pressure-induced changes in the character
of the conduction electrons through \textit{s-p} transfer in Li and Na and
\textit{s-d} transfer in K, Rb, and Cs are believed to destabilize the highly
symmetric low-pressure structures (bcc, fcc) and favor complex, lower symmetry
structures \cite{young}. Another result of the marked change in conduction
electron character is that the superconducting transition temperature of Li
soars from 0.4 mK at ambient pressure to 14 K at 30 GPa \cite{shanti}, an
increase by over four orders of magnitude.

Superconductivity in Cs near 1.3 K was reported many years ago by Wittig
\cite{wittig1} for pressures near 12 GPa, close to the pressure where Cs enters
the orthorhombic \textit{oC}16 phase \cite{schwarz2}. In the Supplemental
Material \cite{sm} very recent experiments are presented that confirm these early
results. It would seem likely that all alkali metals will become
superconducting under pressure, yet no superconductivity has been found for
the following: \ Na in the pressure/temperature ranges to 58 GPa above 4 K
\cite{schilling}; K to 17 GPa above 1.5 K or to 44 GPa above 4 K
\cite{schilling}, as well as to 94 GPa above 1.35 K \cite{debessai}; Rb to 21
GPa above 50 mK \cite{ullrich2}. Schwarz \textit{et al.} \cite{schwarz} have
suggested that a search for Rb should be carried out for pressures of 50 GPa
and above since this is the pressure where Rb takes on the same \textit{oC}16
phase where Cs becomes superconducting.

In the present electrical resistivity studies on Rb the pressure range of
previous work has been extended to $\sim80$ GPa for temperatures above 1.3 K.
Strong evidence is given for the appearance of superconductivity in Rb near 2
K following a sluggish phase transition at 50 GPa from the\ tetragonal
\textit{tI}4 to the orthorhombic \textit{oC}16 phase. Rb thus becomes the 55th
elemental superconductor in the periodic table.

Polycrystalline Cs (99.98 \% pure) and Rb (99.75 \% pure) samples were
obtained in sealed glass ampules from Alfa Aesar. A diamond anvil cell (DAC)
made of conventional and binary CuBe \cite{schilling1} was used to generate
pressures to $\sim80$ GPa between two opposed diamond anvils (1/6-carat, type
Ia) with 0.5 mm diameter culets. The force applied to the anvils was generated
by a stainless-steel diaphragm filled with He gas \cite{daniels}. The Re
gasket (250 $\mu$m thick) was preindented to 80 $\mu$m and a 250 $\mu$m
diameter hole drilled through the center of the preindentation area. A
cBN-epoxy insulation layer was compressed onto the surface of the gasket (see
Fig 1). Four Pt strips (4 $\mu$m\ thick) were then placed on the insulation
layer, acting as electrical leads for the four-point resistivity measurement.
Further details are published elsewhere \cite{shimizu,lim}.

Whereas a small, thin lanthanide sample was normally placed on top of the four
Pt strips (see Fig 1 in Ref \cite{lim}), the extreme softness of the alkali
metals led to large sample extrusion out of the cell, necessitating shaping a
bowl in the cBN-epoxy cell to contain the alkalis, as illustrated in Fig 1.
Because of their extreme air sensitivity, the Cs and Rb samples were loaded
into the bowl-shaped chamber inside a glove box filled with ultra-high purity
Ar gas.

\begin{figure}[t]
\includegraphics[width = 8 cm]{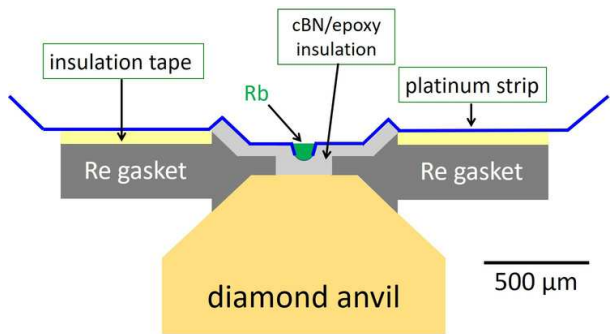}
\caption{\label{fig1}(color online) Schematic drawing of pressure cell with Re gasket used in
the present four-point electrical resistivity experiments on pure Cs and Rb.
Unlike the standard configuration (see Fig 1 in Ref \cite{lim}), here a bowl
is formed in the central cBN/epoxy section to contain the soft alkalis. The
opposing diamond anvil presses downward from above generating pressures to
$\sim80$ GPa.}
\end{figure}

The DAC was inserted into an Oxford flow cryostat capable of varying
temperature from ambient to 1.3 K. Pressure was determined at room temperature
using the diamond vibron \cite{akahama}. With a ruby manometer the pressure
can normally be measured \textit{in situ} over the entire temperature range
\cite{chijioke}. Unfortunately, the ruby sphere would normally sink into the
soft alkali metal, the \textit{R}1 fluoresence being lost. However, in one
experiment to 14 GPa on Cs the ruby fluorecence could be detected to low
temperatures, revealing a pressure increase of $\sim17\%$ on cooling from 295
to 4 K. This pressure increase was used in experiments on both Cs and Rb to
estimate the pressure at low temperatures from the measured vibron pressure at
ambient temperature.

In Fig 2 the resistance of Rb at ambient temperature (295 K) is plotted versus
both increasing and decreasing pressure. Initially the pressure was slowly
increased along the black line to 73 GPa, then decreased to 34 GPa, followed
by an increase along the red line to 75 GPa, then decreased to 48 GPa. The
many structural phase transitions that occur in pure Rb over this pressure
range are shown at the top of the graph \cite{nelmes}. A small peak in
\textit{R(P)} marks the bcc $\rightarrow$ fcc transition 7 GPa with a much
more prominent structure in \textit{R(P)} for the next transitions. In the
tetragonal \textit{tI}4 phase the resistance changes little with pressure, but
begins to increase sharply after passing the \textit{tI}4 $\rightarrow$
\textit{oC}16 phase boundary at 48 GPa. The sizable hysteresis in
\textit{R(P)} reveals the first-order nature of this transition that is
reportedly very sluggish at ambient temperature \cite{schwarz}. The
\textit{R(P)} dependence and structural phase diagram at ambient temperature
for Cs bear a marked similarity to those for Rb in Fig 2 if the differences in
the transition pressures are taken into account, as seen in the Supplemental Material
\cite{sm}.

\begin{figure}[t]
\includegraphics[width = 8 cm]{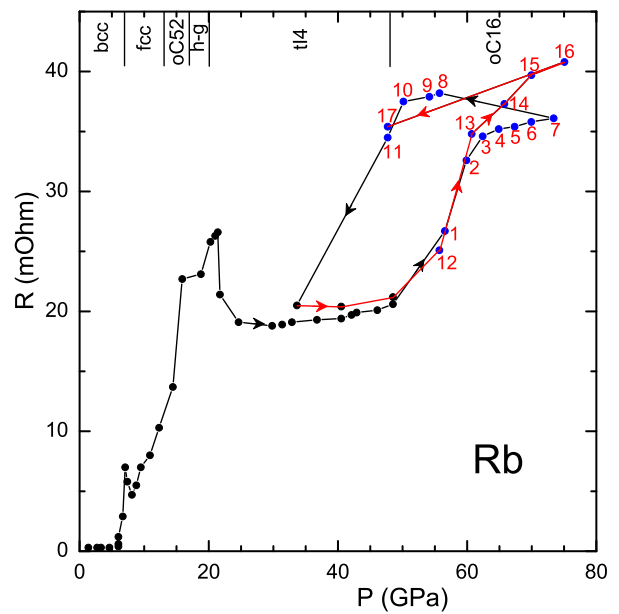}
\caption{\label{fig2}(color online) Dependence of room-temperature resistance of Rb on pressure
to 75 GPa. Arrows give direction of pressure change. Beginning at lowest
pressure, data are taken along black line to point 7, followed by pressure
decrease to 34 GPa. The red line then tracks second increase in pressure to
point 16 followed by decrease to point 17. Beginning with data point 1 on
black line, all further data points \textit{R(T)} were measured between 295
and 1.3 K. For blue data points superconductivity was observed near 2 K; for
black data points no evidence for superconductivity was found above 1.3 K.
Numbers give order of measurement for blue data points first on the black
line, then on the red line. Crystal structures at top of graph from Ref
\cite{nelmes}.}
\end{figure}

In Fig 3 the temperature-dependent resistance of Rb is displayed for 61, 75,
and 48 GPa, measured in that order, as seen by comparing the order of
measurement numbers to those in Fig 2. At all three pressures a sharp drop in
the resistance is observed near 2 K (see inset in Fig 3 for 56 GPa), as would
be expected for a superconducting transition. As discussed below, it is common
in superconducting alkali metals that the resistivity fails to disappear
completely in high-pressure experiments. A marked negative curvature is
evident in the temperature dependence of the resistance, a hallmark of
transition metals. This is clear evidence for strong \textit{d}-character in
the conduction electrons of Rb at these very high pressures arising from
\textit{s-d} electron transfer.

\begin{figure}[t]
\includegraphics[width = 8 cm]{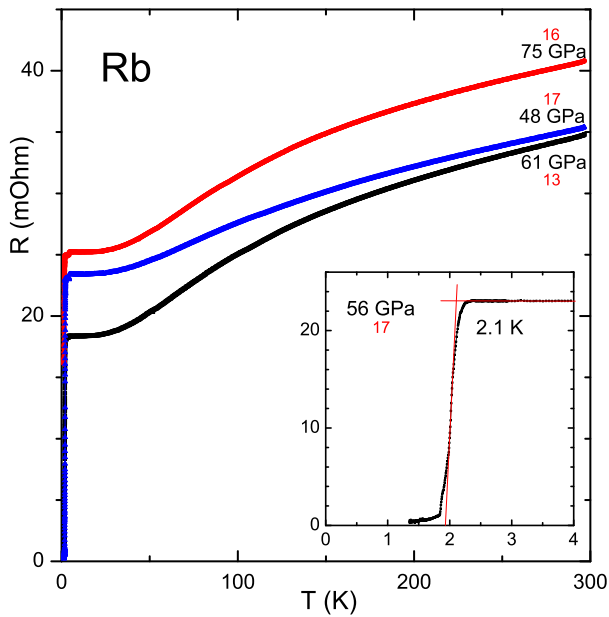}
\caption{\label{fig3}(color online) Resistance of Rb versus temperature at three different
pressures where superconducting transition is seen at low temperature. Red
numbers give order of measurement from Fig 2. Inset shows transition at 2.1 K
for 56 GPa. Value of transition temperature is determined by intersection
point of two straight lines.}
\end{figure}

In Fig 4 a sharp drop in the resistance is also observed for Rb at 71 GPa. As
the magnet field is increased to 480 G, the sharp drop is seen to shift to
progressively lower temperatures at the rate $\sim1$ mK/G. The sharp
resistance drop plus the shift of the transition to lower temperatures in a
magnetic field give strong evidence that Rb metal becomes superconducting for
pressures above 56 GPa.

\begin{figure}[t]
\includegraphics[width = 8 cm]{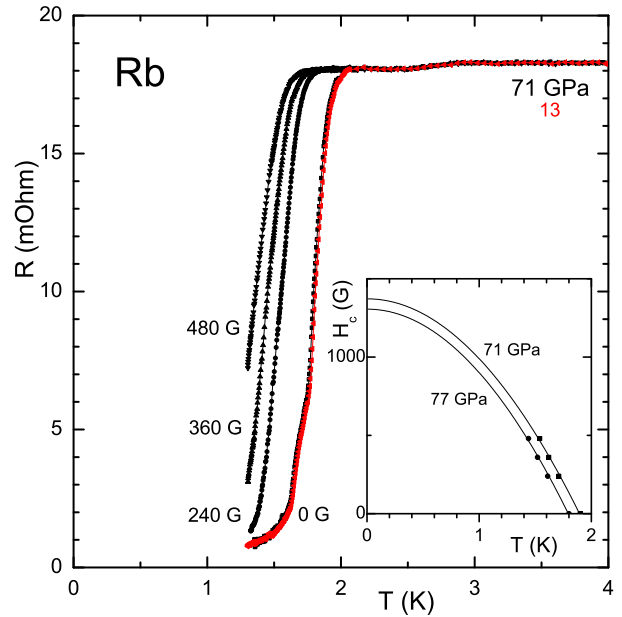}
\caption{\label{fig4}(color online) Resistance Rb versus temperature to 4 K at 71 GPa for four
different values of magnetic field 0, 240, 360, and 480 G. Superconducting
transition shifts to lower temperatures with increasing magnetic field. Inset
shows dependence of transition temperature on magnetic field at 71 and 77 GPa
pressure. Fit of data allows estimate of critical field $H_{0}$ at 0 K (see
text). Number "13" gives order of data point from Fig 2.}
\end{figure}

In the inset in Fig 4 the dependence of the critical temperature on magnetic
field is plotted at both 71 and 77 GPa. From a fit to these data using the
standard expression $H_{\text{c}}(T)=H_{0}[1-(T/T_{\text{c}})^{2}]$
\cite{ashcroft3}, the critical field at 0 K, $H_{0},$ can be estimated for Rb
to be $H_{0}$ $\approx$ 1370 and 1310 G at 71 and 77 GPa, respectively. In the
Supplemental Material \cite{sm} it is pointed out that for Cs $H_{0}$ $\approx$ 270 G at
15.1 GPa. For Li from Ref \cite{shanti} $H_{0}$ $\approx$ 800 and 1000 G at 22
and 24 GPa, respectively.

In Fig 5 the dependence of the resistance of Rb on temperature is plotted to 3
K at various pressures. Except for the data at 56 GPa, the transition
temperature $T_{\text{c}}$ is seen to decrease monotonically with increasing
pressure. Interestingly, the resistance at 3 K shows a non-monotonic pressure
dependence that reflects that shown in Fig 2 for the resistance at room
temperature. The dependences of \textit{T}$_{\text{c}}$ on pressure from Fig 5
and a second experiment are shown in Fig 6, yielding from the straight-line
fit the value $dT_{\text{c}}/dP\simeq-39$ mK/GPa.

\begin{figure}[t]
\includegraphics[width = 8 cm]{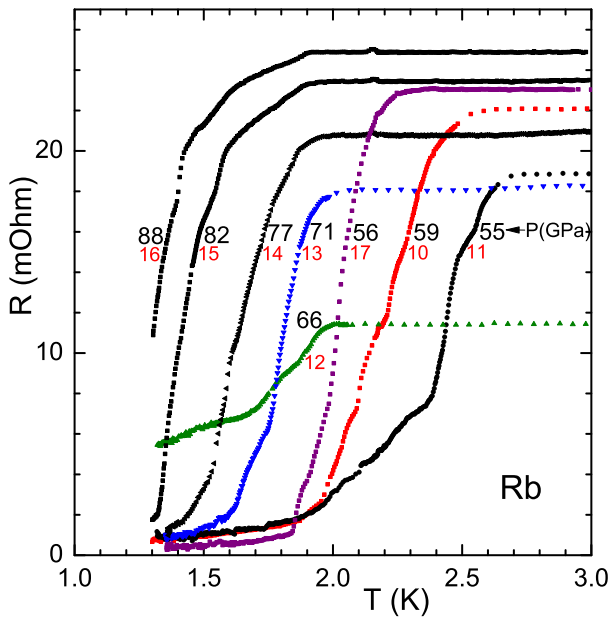}
\caption{\label{fig5}(color online) Resistance of Rb versus temperature to 3 K showing
superconducting transitions at various pressures. Black numbers give value of
pressure for each transition. Red numbers give order of measurement from Fig
2. Smaller size of transition at 66 GPa gives evidence sample is in mixed phase.}
\end{figure}

\begin{figure}[t]
\includegraphics[width = 8 cm]{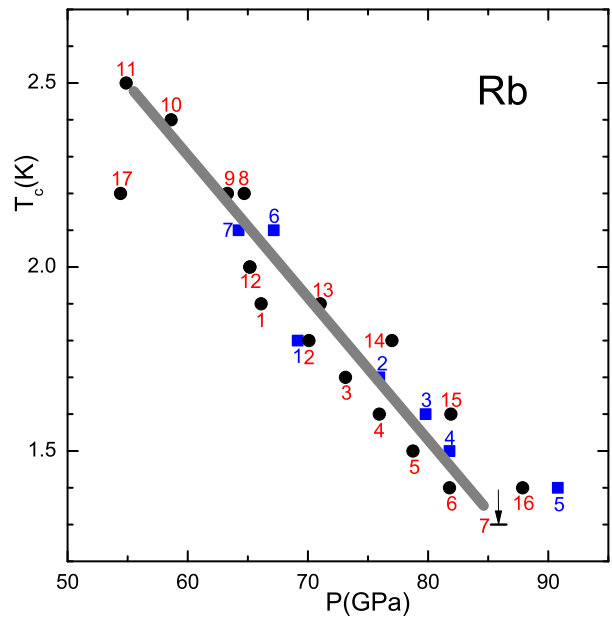}
\caption{\label{fig6}(color online) Dependence of transition temperature $T_{\text{c}}$ of Rb
on pressure to 88 GPa in two separate experiments (black $\bullet$, blue
$\blacksquare$). Broad grey straight line gives rough fit to data with slope
$dT_{\text{c}}/dP\simeq-39$ mK/GPa.}
\end{figure}

From the above results it is clear that superconductivity in Rb above 1.3 K
only occurs in its \textit{oC}16 phase, but not in the \textit{tI}4 phase. As
this sluggish phase transition begins with increasing pressure, the
resistivity starts to rise, as seen in Fig 2. At 48 GPa the sample is still in
the \textit{tI}4 structure and no superconductivity is observed. However, upon
increasing pressure to 56 GPa (points 1 or 12) the resistance has clearly
increased indicating that at least some of the \textit{tI}4 phase has
transitioned to \textit{oC}16. That only a partial superconducting transition
is seen for points 1 and 12 (see Fig 5 for point 12) shows that at 56 GPa the
sample is still in a mixed phase region. As the pressure is increased further
to 75 GPa (points 12-16 or 1-7 in Fig 2), the resistivity begins to saturate,
possibly indicating the phase transition to \textit{oC}16 is nearing
completion. At the same time the superconducting transition becomes more
complete. At no pressure does the resistance get closer than 98\% to zero.
This effect of non-zero resistance observed here for Rb as well as for Cs in the
Supplemental Material \cite{sm} and earlier by Ullrich
\cite{ullrich,wittig}, arises from the non-ideal geometry in the resistivity
experiment and the inhomogeneous pressure distribution across the sample.

Since the superconducting transition temperature $T_{\text{c}}$ for Rb
decreases monotonically with increasing pressure, the highest value of
$T_{\text{c}}$ would occur at the lowest pressure, as long as the sample
remains in the high-pressure \textit{oC}16 phase. These facts were used to
maximize the value of $T_{\text{c}}$ by decreasing pressure from point 16 to
17 (or from point 7 to points 10 or 11) in Fig 2 where Rb is still mainly in
the \textit{oC}16 phase on unloading due to the large $\sim15$ GPa hysteresis
in \textit{R(P)}. Indeed, the $R(T)$ data in Fig 5 for points 10, 11, 17 place
$T_{\text{c}}$ in the range 2.2 - 2.5 K, the highest values measured in the
present experiment.

The importance of the \textit{oC}16 phase for the appearance of
superconductivity above 1.3 K is not restricted to Rb but also applies to
neighboring Cs, as emphasized by Schwarz \textit{et al.} \cite{schwarz}. Once
in the \textit{oC}16 phase, the transition temperature $T_{\text{c}}$ for both
Rb and Cs is observed to decrease with increasing pressure. In addition, for
Cs the transition between the same two phases \textit{tI}4 $\rightarrow$
\textit{oC}16 is accompanied by a marked increase in the resistivity, as seen in the 
Supplemental Material \cite{sm}. In electronic structure calculations on Rb
by Fabbris \textit{et al.} \cite{fabbris} a prominent peak structure already appears in
the density of states near the Fermi level for pressure near 16 GPa, the
conduction electrons taking on a significant \textit{d}-character. At the fcc
$\rightarrow$ \textit{oC}52 phase transition a sizable enhancement in the density of states occurs.
Perhaps the sudden appearance of superconductivity in Rb near 55 GPa is the
result of a similar enhancement at the \textit{tI}4 $\rightarrow$ \textit{oC}16 phase transition.

In the next lighter alkali metal K the \textit{tI}4 $\rightarrow$
\textit{oC}16 transition also occurs, but not until the much higher pressure
of 96 GPa \cite{lunde}. M. Debessai \cite{debessai} found no superconductivity
in K above 1.35 K to 94 GPa pressure. A search for superconductivity in K to
pressures higher than 1 Mbar would thus seem warranted.

In all three heavy alkali metals Cs, Rb, and K the importance of the
pressure-induced \textit{s-d} electron transfer has been emphasized. It is
likely responsible for the transition from tetragonal \textit{tI}4 to the
lower symmetry orthorhombic \textit{oC}16 structure. McMahan \cite{mcmahan}
has calculated the completion pressure for the \textit{s-d} transfer in Cs and
Rb to be 15 and 53 GPa, respectively, near the pressures where the
\textit{tI}4 $\rightarrow$ \textit{oC}16 transition in both cases begins, but
also where the resistivity rises steeply and superconductivity appears. A
parallel calculation \cite{mcmahan} for K finds its \textit{s-d} transfer to
complete at $\sim60$ GPa, a lower pressure than that for the \textit{tI}4
$\rightarrow$ \textit{oC}16 transition.

In summary, a superconducting transition $T_{\text{c}}$ near 2 K has been
observed in four-point electrical resistivity measurements on Rb for pressures
above 55 GPa. The transition only occurs after Rb enters the orthorhombic
\textit{oC}16 phase which is marked by a sharp increase in the resistivity.
$T_{\text{c}}$ decreases under pressure at the rate $-39$ mK/GPa. In all
respects these results parallel those found for Cs near 12 GPa pressure. Based
on these results it is predicted that K will become superconducting near 1
Mbar pressure when it enters the same \textit{oC}16 phase.

\noindent\textbf{Acknowledgments.} The authors would like to thank G. Fabbris
for hepful comments on this manuscript. This research is supported by the
National Science Foundation (NSF) through Grants No. DMR-1104742 and No.
DMR-1505345 as well as by the Carnegie/DOE Alliance Center (CDAC) through
NNSA/DOE Grant No. DE-FC52-08NA28554.

\label{biblio}


\begin{thebibliography}{99}                                                                                               %


\bibitem {tuoriniemi}J. Tuoriniemi, K. Juntunen-Nurmilaukas, J. Uusvuori, E.
Pentti, A. Salmela, and A. Sebedash, Nature (London) \textbf{447}, 187 (2007).

\bibitem {schilling}J. S. Schilling, High Press. Research \textbf{26}, 145 (2006).

\bibitem {takemura}K. Takemura, N. E. Christensen, D. L. Novikov, K. Syassen,
U. Schwarz and M. Hanfland, Phys. Rev. B \textbf{61}, 14399 (2000).

\bibitem {neaton}J. B. Neaton and N. W. Ashcroft, Nature \textbf{400}, 141
(1999); J. Feng, R. G. Hennig, N. W. Ashcroft, and R. Hoffmann, Nature
\textbf{451}, 445 (2008); P. S\"{o}derlind, O. Erikson, B. Johansson, J. M.
Wills, and A. M. Boring, Nature (London) \textbf{374}, 524 (1995).

\bibitem {dye}J. L. Dye, Science \textbf{301}, 607 (2003).

\bibitem {young}D. A. Young, \textit{Phase Diagrams of the Elements}
(University of California Press, Berkeley, 1991).

\bibitem {shanti}S. Deemyad and J. S. Schilling, Phys. Rev. Lett. \textbf{91},
167001 (2003).

\bibitem {wittig1}J. Wittig, Phys. Rev. Lett. \textbf{24}, 812 (1970).

\bibitem {schwarz2}U. Schwarz, K. Takemura, M. Hanfland, and K. Syassen, Phys.
Rev. Lett. \textbf{81}, 2711 (1998).

\bibitem {sm}See Supplemental Material at http:// \ for results of present
high-pressure resistivity studies on Cs metal, which includes Refs.
\cite{mcmahon,ma}.

\bibitem {mcmahon}M. I. McMahon, R. J. Nelmes, and S. Rakhi, Phys. Rev. Lett.
\textbf{87}, 255502 (2001).

\bibitem {ma}Y. Ma, M. Eremets, A. R. Oganov, Y. Xie, I. Trojan, S. Medvedev,
A. O. Lynkhov, M. Valle, and V. Prakapenka, Nature (London) \textbf{458}, 182 (2009).

\bibitem {debessai}M. T. Debessai, Ph.D. thesis, Washington University in St
Louis (2004).

\bibitem {ullrich2}K. Ullrich, C. Probst and J. Wittig, J. de Physique,
Colloque \textbf{6}, 463 (1978).

\bibitem {schwarz}U. Schwarz, K. Syassen, A. Grzechnik, and M. Hanfland, Solid
State Commun. \textbf{112}, 319 (1999).

\bibitem {schilling1}J. S. Schilling, in \textit{Proceedings of the 9th AIRAPT
International High Pressure Conference}, Albany, New York, July 24-29, 1983,
edited by C. Homan, R. K. MacCrone and E. Whalley (North-Holland, New York,
1984); J. S. Schilling, Mater. Res. Soc. Symp. Proc. \textbf{22}, 79 (1984).

\bibitem {daniels}W. B. Daniels and W. Ryschkewitsch, Rev. Sci. Instrum.
\textbf{54}, 115 (1983).

\bibitem {shimizu}K. Shimizu, K. Amaya, and N. Suzuki, J. phys. Soc. Jpn.
\textbf{74}, 1345 (2005).

\bibitem {lim}J. Lim, G. Fabbris, D. Haskel, and J. S. Schilling, Phys. Rev. B
\textbf{91}, 045116 (2015).

\bibitem {akahama}Y. Akahama and H. Kawamura, J. Appl. Phys. \textbf{100},
043516 (2006).

\bibitem {chijioke}A. D. Chijioke, W. J. Nellis, A. Soldatov, and I. F.
Silvera, J. Appl. Phys. \textbf{98}, 114905 (2005).

\bibitem {nelmes}R. J. Nelmes, M. I. McMahon, J. S. Loveday, and S. Rekhi,
Phys. Rev. Lett. \textbf{88}, 155503 (2002).

\bibitem {ashcroft3}N. W. Ashcroft and N. D. Mermin, in Solid State Physics
(Thomson Learning, Inc. 1976) p.745.

\bibitem {ullrich}K. Ullrich, Ph.D. thesis, University of Cologne (1980).

\bibitem {wittig}J. Wittig, in \textit{Superconductivity in d- and f-Band
Metals}, edited by W. Buckel and W. Weber (Kernforschungszentrum, Karlsruhe,
1982) p. 321.

\bibitem {fabbris}G. Fabbris, J. Lim, L. S. I. Veiga, D. Haskel, and J. S.
Schilling, Phys. Rev. B \textbf{91}, 085111 (2015).

\bibitem {lunde}L. F. Lundegaard, M. Marques, G. Stinton, G. J. Ackland, R. J.
Nelmes, M. I. McMahon, Phys. Rev. B \textbf{80}, 020101 (2009).

\bibitem {mcmahan}A. K. McMahan, Phys. Rev. B \textbf{29}, 5982 (1984).
\end{thebibliography}
\end{document}